# A coupled planar transmit RF array for ultrahigh field spine MR imaging

Yunkun Zhao, Student Member, IEEE, Komlan Payne, Graduate Student Member, IEEE, Leslie Ying, Senior Member, IEEE, and Xiaoliang Zhang, Member, IEEE

*Abstract*— Ultrahigh-field MRI, such as those operating at 7 Tesla, enhances diagnostic capabilities but also presents unique challenges, including the need for advanced RF coil designs to achieve an optimal signal-to-noise ratio and transmit efficiency, particularly when imaging large samples. In this work, we introduce the coupled planar array, a novel technique for high-frequency, large-size RF coil design with enhanced the RF magnetic field (B1) efficiency and transmit performance for ultrahigh-field spine imaging applications. This array comprises multiple resonators that are electromagnetically coupled to function as a single multimodal resonator. The field distribution of its highest frequency mode is suitable for spine imaging applications. Based on the numerical modeling and calculation, a prototype of the coupled planar array was constructed and its performance was evaluated through comprehensive numerical simulations, rigorous RF measurements, empirical tests, and a comparison against a conventional surface coil with the same size and geometry. The results of this study demonstrate that the proposed coupled planar array exhibits superior performance compared to conventional surface coils in terms of B1 efficiency for both transmit (B1+) and receive (B1-) fields, specific absorption rate (SAR), and the ability to operate at high frequencies. This study suggests a promising and efficient approach to the design of high-frequency, large-size RF coils for spine MR imaging at ultrahigh magnetic fields.

*Index Terms*—Ultra-high field, RF Coil, Surface Coil, B$_1$ Efficiency, Specific Absorption Rate, MR imaging

## I. INTRODUCTION

MAGNETIC Resonance Imaging (MRI) is a pivotal technology for visualizing soft tissues and determining metabolic processes, providing high-resolution images with varied contrasts without employing ionizing radiation [1, 2]. Advancements in MRI technology have led to the development of high and ultrahigh field systems, notably the 7 Tesla MRI, which is now increasingly utilized in clinical settings [3-9]. The transition to a higher static magnetic field strength, from the standard 1.5T or 3T to 7T, significantly enhances the signal-to-noise ratio (SNR) [10-14]. This improvement not only boosts image resolution and spectral dispersion, but also reduces scan times, utilizing the parallel imaging [15-17] and compressed sensing based fast imaging techniques[18]. Additionally, ultrahigh field MRI enhances vasculature conspicuity, improves angiography, and augments spectroscopy acquisitions[19-22].

Despite the substantial benefits of ultrahigh field MRI, the required high resonance frequency of the ultrahigh field, e.g. ~300MHz at 7T, poses significant challenges in designing large transmit RF coils for MR signal excitation, which can impede its optimal performance and functionality [23-28]. The key requirements in designing large transmit RF coils in ultrahigh field MRI are (1) high efficiency of transmit RF magnetic field (B1+), (2) ability to achieve high resonant frequency, and (3) low specific absorption rate (SAR). Large size RF coils, e.g. large L/C loops, inherently face difficulties in achieving high transmit efficiency due to their physical dimensions and the circuit layout . Due to their intrinsically high inductance, the large size transmit coils suffer from achieving the required high resonance frequency. To overcome the inductance challenges and operate at the required high frequency, the capacitance of large size transmit coils must be significantly reduced. However, reducing capacitance can generate high electric fields, potentially increasing the specific absorption rate (SAR) and tissue heating, ultimately posing safety hazards to patients or subjects being imaged. Due to these technical challenges, currently no large size transmit RF coils are available in clinical ultrahigh field MR systems.

In recent years, methods using multichannel transmit or transceive arrays to excite large samples, e.g. the spine, at the ultrahigh field of 7T have been explored [29-31]. High-channel-count phased array coils, such as 8-channel transmit arrays, have demonstrated enlarged excitation coverage and multichannel transmit capability compared to earlier single-channel designs for cervical spinal cord imaging at 7T [30, 31]. However, imaging with multichannel transmit arrays requires high-power multichannel transmitters, which are not cost-effective and are often unavailable in standard MR scanners. Developing a technique for designing large, single-channel, and highly efficient transmit coils would be highly beneficial for ultrahigh field MR imaging applications.

To address the challenges encountered in the design of large transmit coils at ultrahigh fields, in this work, we propose and

This work was supported in part by the NIH under Grant U01 EB023829 and SUNY Empire Innovation Professorship. (Corresponding author: Xiaoliang Zhang.)

Yunkun Zhao and Komlan Payne are with Department of Biomedical Engineering, State University of New York at Buffalo, Buffalo, NY 14260 USA (e-mail: yunkunzh@buffalo.edu).

Leslie Ying is with the Departments of Biomedical Engineering and Electrical Engineering, State University of New York at Buffalo, Buffalo, NY 14260 USA (e-mail: leiying@buffalo.edu).

Xiaoliang Zhang is with the Departments of Biomedical Engineering and Electrical Engineering, State University of New York at Buffalo, Buffalo, NY 14260 USA (e-mail: xzhang89@buffalo.edu).

investigate a novel solution: the coupled planar RF array for high-frequency large-size transmit coils [32]. This design comprises multiple L/C loop resonators, which are arranged in a planar array with no overlap. All the resonators are electromagnetically coupled, thereby forming a single multimodal resonator. Among those multiple resonant modes, the one resonating at the highest frequency exhibits a distinctive magnetic field distribution, analogous to that of conventional loop surface coils and is therefore suitable for MR imaging. The experimental results demonstrate that the proposed coupled planar array outperforms the conventional surface coils in spine imaging at the ultrahigh field of 7T in terms of high-frequency operation, B1 efficiency and penetration, and SAR. This coupled planar array technique provides a simple and robust solution to designing high-frequency large-size transmit coils for ultrahigh field spine MR imaging with improved transmit efficiency and lower SAR, which ultimately enhances the functionality and patient safety of ultrahigh field MR.

## II. Description of Methodology

### A. EM Simulation

Figure 1 presents the simulation model of the coupled planar array. This array consists of five resonators or coil loops, each constructed from 16 AWG copper wire and sequentially arranged into a coil array. Four of these coils are identical square units, each measuring 10 cm on each side and equipped with four 4.1pF capacitance tuning capacitors to tune the resonators to 263 MHz. When electromagnetically coupled, the array operates at a higher resonant frequency of 300 MHz, which is achieved through the combined effects of the coupled design. The central coil includes a driving port and an impedance matching circuit, crucial for the operation of the coupled planar array. The coils are spaced only 1 mm apart to maximize mutual inductive coupling, with the total length of the array extending to 50 cm and a width of 10 cm. Designed with multiple coils, our array supports three resonant modes. We utilize the highest resonant mode for imaging applications, chosen for its ability to produce a uniform B1 field direction and reach high frequencies effectively. In this study, we evaluated the performance of our coupled planar array against a large, single-loop conventional surface coil to establish comparative benchmarks. The conventional surface coil, constructed using 16 AWG wire, mirrors the dimensions of our design, measuring 50 cm in length and 10 cm in width. It features a tuning arrangement comprising twelve 2.15 pF capacitors, all meticulously tuned to operate at 300 MHz. Each coil in this array is equipped with three 3.2 pF capacitors, and overlapping between the coils has been implemented as a method for achieving effective decoupling. The total size of the decoupled array is slightly smaller than our coupled array, measuring 49 cm in length, due to the overlapping necessary for decoupling, and 10 cm in width. This setup allows us to directly compare its performance with our coupled planar array under identical operational conditions. In simulation, all designs are placed 1 cm below an oil phantom with a dimension of 70×30×15 cm2. An alternative simulation comparison has also been applied using CST body model Gustav. The bio model has also been placed 1 cm above the coil designs. Performance assessments of the study involved analyzing scattering parameters, SAR, and B1 efficiency, using field distribution plots. All electromagnetic field plots were normalized to 1 W of total accepted power. Numerical results of the proposed designs were obtained using the electromagnetic simulation software CST Studio Suite (Dassault Systèmes, Paris, France).

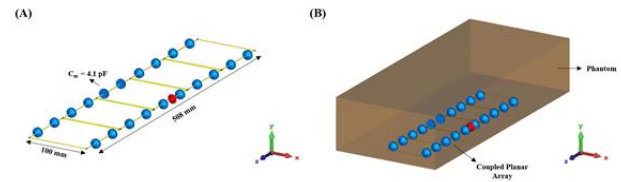

Fig. 1. (A) Simulation model of the proposed coupled planar array, showing the dimensions and detailed structure. (B) Proposed coupled planar array loaded with a cuboid phantom.

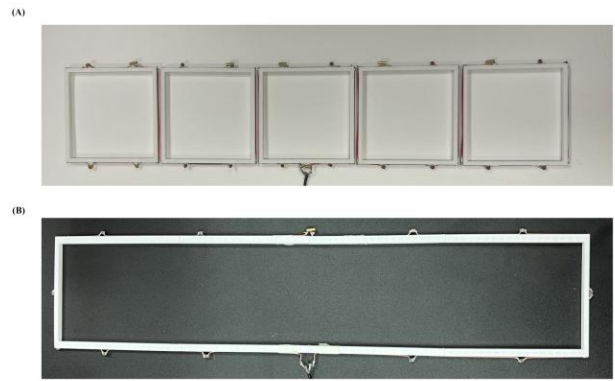

Fig. 2. Bench test model of (A) coupled planar array and (B) large, single-loop conventional surface coil.

### B. Bench Test Model Assembly

Figure 2 presents photographs alongside dimensional details of the bench test models for our coupled planar array and the comparative designs. These test models retain the same dimensions as specified in the simulation models, ensuring consistency across experimental and simulated setups. The coupled planar array was constructed with 16 AWG copper wire and built upon a 3D printed polylactide structure fabricated using a Flashforge Guider 2s 3D printer (Flashforge, Zhejiang, China). The array was meticulously tuned to a resonant frequency of 298 MHz, aimed at matching the operational frequency of 7T ultrahigh field MRI systems. The array incorporates twenty identical 4pF capacitors, ensuring uniformity and reliability in performance.

A large, single-loop conventional surface coil was constructed for comparison purposes. The coil was built using 16 AWG copper wire and tuned with eleven 1.8 pF capacitors. To ensure structural stability and maintain consistency with the coupled planar array, the coil was also mounted on a 3D-printed polylactide structure. This setup facilitated a direct and reliable performance comparison with the proposed coupled planar array.

The experimental configuration incorporates an H-field sniffer, which is mounted on a high-precision router system, the Genmitsu CNC PROVerXL 4030. This setup facilitates the precise placement of the field probe to capture the B1 field emitted by the RF coil in a three-dimensional space. This probe is connected to a Keysight E5061B vector network analyzer (Santa Clara, CA, U.S), which collects the raw data including output, accepted power, and scattering parameters. The data captured by the vector network analyzer is then transmitted to a computer and analyzed using MATLAB to generate the B1 field efficiency map. The efficiency mapping is conducted on a 30 x 7 cm slice in the Y-Z plane, positioned 1 cm above the RF coils, with measurements taken every 2.5 mm. All experimental results are normalized to 1 watt of accepted power to standardize the output across tests.

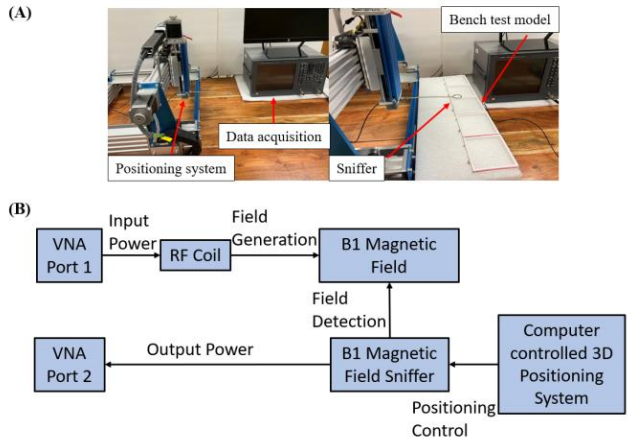

Fig. 3.  (A) Photograph of measurement setup including the sniffer-positioning system, network analyzer and the data process computer. (B) Schematic of the measurement workflow to obtain B1 field efficiency map.

## III. RESULT

### A. Simulated Resonant Frequency and Field Distribution

Figure 4 illustrates the simulated scattering parameters vs frequency for the coupled planar array. The figure displays three distinct split resonant peaks at 238 MHz, 261 MHz, and 298 MHz, with the highest peak at 298 MHz designated for imaging applications. Notably, each coil within the coupled planar array resonates naturally at a lower frequency of 263 MHz, highlighting the success of our tuning strategy that utilizes a higher capacitance value to achieve increased resonant frequencies for coils of equivalent size. In Figure 5A, the B1 field efficiency maps are shown across the Y-Z and X-Z planes inside the phantom. These maps depict the efficiency of the B1 field generated by the coupled planar array, demonstrating its capability to achieve enhanced B field efficiency with a B1 field distribution on par with conventional

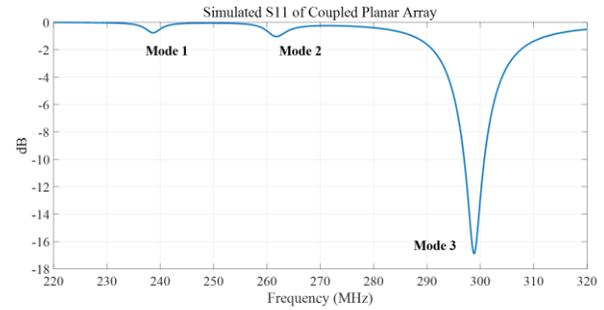

Fig. 4.  Simulated reflection coefficient S11 vs. frequency of the coupled planar array. Three resonant modes with frequencies of 238 MHz, 261 MHz, and 298 MHz were observed.

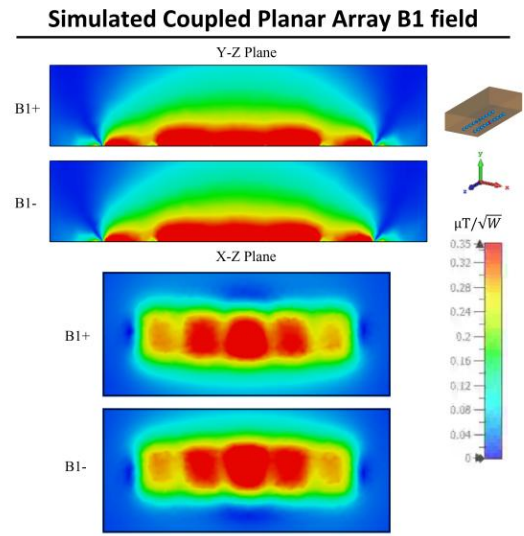

Fig. 5.  (A) Simulated Y-Z and X-Z plane B field efficiency maps inside phantom generated by coupled planar array.

surface coil.

Figure 6 depicts the simulated distribution of surface currents across each coil layer. It is observed that the currents flow in a uniform counterclockwise direction throughout the coils, reflecting a consistent and effective driving field. An increment in the magnitude of surface current is particularly noticeable in the central driving coil (Coil 3), decreasing progressively toward the peripheral layers. This pattern of current intensity can be attributed to the electromagnetic coupling within the coils, with the central coil directly connected to the power source, inducing more robust currents.

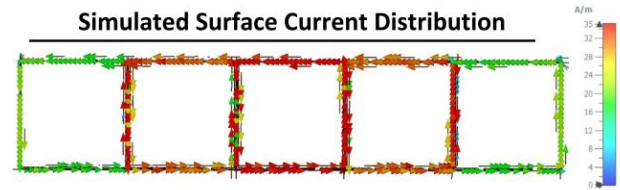

Fig. 6.  Simulated surface current distribution plot of the multimodal surface coil. Coil 1 is positioned at the left rear side, with each subsequent coil (Coil 2 through Coil 5) positioned progressively to the right, culminating in Coil 5 at the rear right. Coil 3 is designated as the driving coil.

As distance from the center increases, the influence of the driving coil diminishes, a result consistent with the increasing impedance and potential energy losses encountered as the current moves through the layered resonant structure.

### B. Measured Scattering Parameters and Field Distribution

Figure 7 displays the S-parameter versus frequency plots for the coupled stack-up coil, demonstrating a close alignment with the predicted simulation outcomes. This figure reveals the formation of three resonant modes at frequencies of 244.8 MHz, 266.6 MHz, and 300.6 MHz. Figure 8 illustrates the B1 field efficiency distribution map on the Y-Z planes, obtained using a 3-D magnetic field mapping system. The coupled planar array exhibits robust B field efficiency and maintains a consistent field distribution pattern across the Y-Z planes, corroborating the simulation results. This congruence underlines the precision and reliability of the simulations, confirming the effectiveness of the coupled planar array in achieving expected performance metrics.

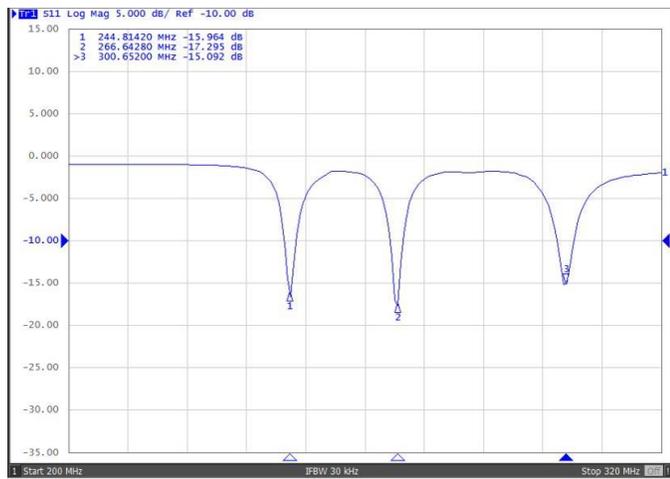

Fig. 7. S11 reflection measurement vs. frequency of the bench test model of coupled planar array.

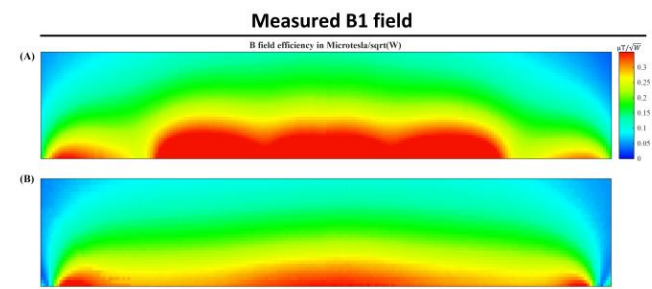

Fig. 8 Measured B field efficiency maps on the Y-Z plane of (A) multimodal surface coil and (B) large-size conventional surface coil.

### C. Field Distribution and Efficiency Evaluation

Figure 9 offers a detailed comparative analysis of the simulated B1 field efficiency between the coupled planar array and a conventional large-size surface coil. The findings clearly illustrate that the coupled planar array achieves a significantly higher B1 field efficiency compared to its conventional counterpart. In Figure 10, a one-dimensional plot of the B1 field efficiency is presented along a horizontal line positioned 3.5 cm and 5 cm above the coils. This visualization offers an in-depth examination of the spatial distribution of the B1 field efficiency. It is evident from the plot that the average B1 field efficiency achieved by the coupled planar array markedly exceeds that of the conventional surface coil. Such results underscore the advanced capabilities of the coupled planar array in uniformly distributing the B1 field across a given area, which is essential for achieving consistent and high-quality imaging results.

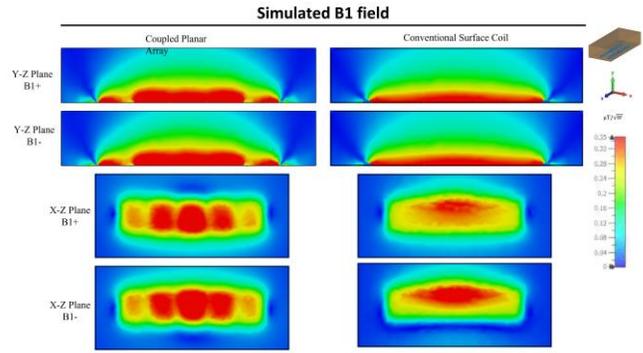

Fig. 9 Simulated Y-Z and X-Z plane B field efficiency maps inside phantom generated by coupled planar array, and conventional surface coil. X-Z planes are at the center of the coil and X-Z plane is 3.5 cm above the coil.

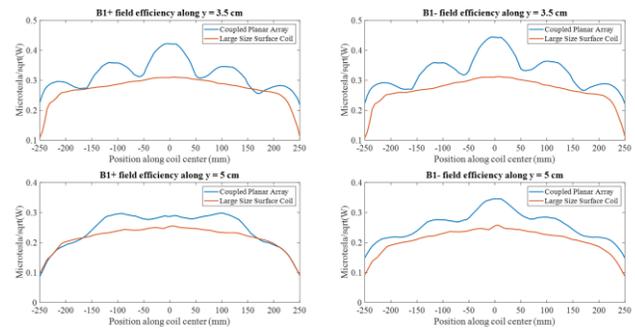

Fig. 10 1-D plot of B1 field efficiency along the horizontal line y = 3.5 cm and y = 5 cm.

Figure 11 depicts a comparative analysis of the B field efficiency between the simulation model of the coupled planar array and the conventional surface coil, each loaded with the CST human bio model named Gustav. Operating at 300 MHz, the B1 field distributions from both coils exhibit resilience against variations in load, maintaining a level of consistency that mirrors the results obtained when an oil phantom is used within the coils. Notably, the B1 field efficiency within the human phantom for the coupled planar array remains significantly higher compared to that of the conventional surface coil. The coupled planar array effectively sustains high B1 field efficiency, as demonstrated through simulations using an oil phantom for mimicking unloaded conditions and a human bio-model for scenarios representative of potential real imaging applications.

### D. Specific absorption rate

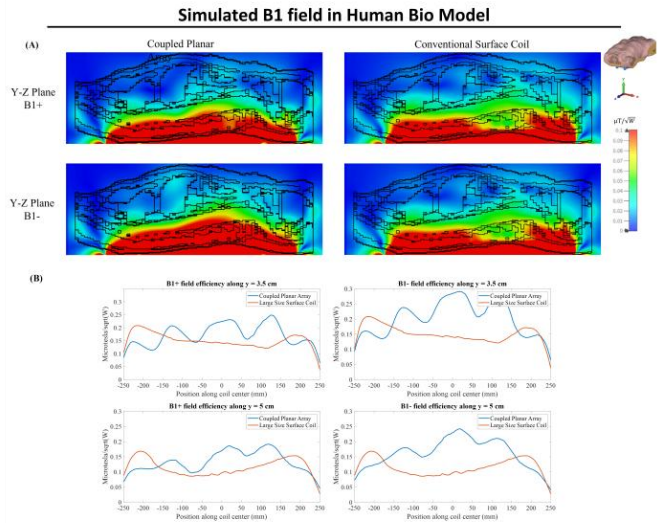

Fig. 11 Comparison between simulated B field efficiency maps generated by coupled planar array and conventional surface coil on the Y-Z plane of the human bio model.

Results of the Specific Absorption Rate (SAR) induced by the coupled planar array and the conventional surface coil are illustrated in Figure 12. This figure demonstrates the simulated SAR values in a human bio model when subjected to both coil types at a frequency of 300 MHz. The coupled planar array shows a peak 10g average SAR of 0.455 W/kg, which is notably lower than the 0.542 W/kg exhibited by the conventional surface coil. The SAR distribution maps clearly highlight the effectiveness of the coupled planar array in minimizing energy absorption by the body, thus enhancing safety. The higher SAR observed on the superior side of the body is possibly attributed to the asymmetric distribution of B1 fields in conductive and high permittivity samples at high frequencies, and also attributed to he natural curvature of the bio-model, which positions the upper side closer to the coil, leading to increased energy deposition in that region.

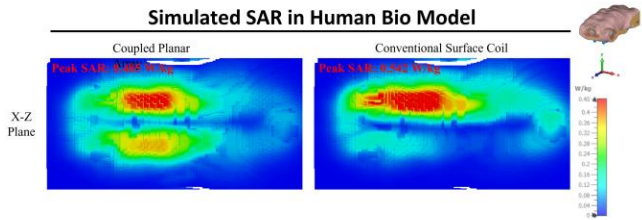

Fig. 12. Comparison between Peak SAR generated by coupled planar array and conventional surface coil on human bio model

## IV. Discussion

One of the primary challenges in designing large transmit coils at ultrahigh fields is achieving the required high frequency while maintaining resonance stability. Reducing capacitance to increase frequency often results in circuit instability due to increased interaction between the resonant structure and its surrounding environment, as well as augmented electric fields. These interactions make the resonance highly sensitive to the environment factors, leading to frequency shifts and, consequently, operational instability or malfunction. While adding multiple splitting capacitors can mitigate some of these issues, it complicates frequency tuning and significantly reduces the tuning range, especially for large coils. The proposed multimodal coupled planar array technique addresses these challenges by leveraging its highest frequency mode, enabling efficient high-frequency operation without the need for excessively small capacitance values. This design ensures excellent resonance stability, high quality factors, and scalability, offering flexibility and reliability for a wide range of clinical and research applications.

High RF power deposition and patient safety are major concerns in ultrahigh field MRI where the required excitation power is significantly higher compared to low field MRI. This safety concern is even more pronounced in large field-of-view (FOV) imaging, where a large transmit coil is typically used and the required excitation power is further increased in order to excite a large volume of sample. The reduced SAR of the proposed multimodal coupled planar array technique is advantageous and can potentially mitigate the tissue heating problem in ultrahigh field imaging.

Improving the efficiency of the transmit B1 fields (B1+) is another way to reduce the excitation power and thus the power deposition in the imaging sample. Due to their large physical size, conventional L/C loop coils suffer from low B1 efficiency and increased excitation power. The multimodal coupled planar array uses multiple small-sized resonators that are electromagnetically coupled to form a large multimodal resonator. This unique structure with more conductors carrying electric currents helps to enhance the B1 fields in the imaging region. Given the circuit structure described in this paper, it is possible to design a double-tuned large-size transmit coil based on the proposed coupled planar array technique to facilitate heteronuclear metabolic MR imaging and spectroscopy at ultrahigh fields.

The scalability of the proposed coupled planar array technique was further evaluated to address concerns regarding the weaker fields observed at the two ends of the array. These weaker fields are possibly due to its intrinsic mode distribution or reduced coupling between the peripheral elements and the central resonators, which inherently limits their current intensity. To mitigate this issue, adjustments such as modifying the spacing or tuning of the end elements could be made to optimize field uniformity across the array. Furthermore, while the current design demonstrates promising scalability for larger excitation volumes, practical considerations, such as maintaining adequate coupling and minimizing field inhomogeneity, must be carefully managed.

## V. Conclusion

In this study, we have successfully developed a high-frequency large-size transmit coil based on the proposed coupled planar array technique for ultrahigh field spine MR imaging. Compared to the conventional loop surface coils with the same size and geometry, the coupled planar array offers superior ability to operate at high frequencies, improved B1 efficiency and penetration, as well as reduced SAR. The coupled planar array technique provides a simple and robust solution to the design of high-frequency large-size transmit coils in spine imaging at ultrahigh fields and ultimately helps enhance the functionality and patient safety of ultrahigh field MRI.

ACKNOWLEDGEMENT

This work is supported in part by the NIH under a BRP grant U01 EB023829 and SUNY Empire Innovation Professorship Award.

REFERENCES

[1] Lauterbur PC. Image Formation by Induced Local Interaction: Examples employing Nuclear Magnetic Resonance. Nature 1973;241:190-1.
[2] Ugurbil K, Garwood M, Ellermann J, Hendrich K, Hinke R, Hu X, et al. Imaging at high magnetic fields: initial experiences at 4 T. Magn Reson Q 1993;9(4):259-77.
[3] Ladd ME, Bachert P, Meyerspeer M, Moser E, Nagel AM, Norris DG, et al. Pros and cons of ultra-high-field MRI/MRS for human application. Progress in Nuclear Magnetic Resonance Spectroscopy 2018;109:1-50.
[4] Barisano G, Sepehrband F, Ma S, Jann K, Cabeen R, Wang DJ, et al. Clinical 7 T MRI: Are we there yet? A review about magnetic resonance imaging at ultra-high field. Br J Radiol 2019;92(1094):20180492.
[5] Wu Z, Zaylor W, Sommer S, Xie D, Zhong X, Liu K, et al. Assessment of ultrashort echo time (UTE) T(2)* mapping at 3T for the whole knee: repeatability, the effects of fat suppression, and knee position. Quant Imaging Med Surg 2023;13(12):7893-909.
[6] Li N, Zheng H, Xu G, Gui T, Yin Q, Chen Q, et al. Simultaneous Head and Spine MR Imaging in Children Using a Dedicated Multichannel Receiver System at 3T. IEEE Trans Biomed Eng 2021;68(12):3659-70.
[7] Opheim G, van der Kolk A, Markenroth Bloch K, Colon AJ, Davis KA, Henry TR, et al. 7T Epilepsy Task Force Consensus Recommendations on the Use of 7T MRI in Clinical Practice. Neurology 2021;96(7):327-41.
[8] Cramer J, Ikuta I, Zhou Y. How to Implement Clinical 7T MRI-Practical Considerations and Experience with Ultra-High-Field MRI. Bioengineering (Basel) 2024;11(12).
[9] de Vries E, Hagbohm C, Ouellette R, Granberg T. Clinical 7 Tesla magnetic resonance imaging: Impact and patient value in neurological disorders. J Intern Med 2025.
[10] Yacoub E, Shmuel A, Pfeuffer J, Van De Moortele PF, Adriany G, Andersen P, et al. Imaging brain function in humans at 7 Tesla. Magn Reson Med 2001;45(4):588-94.
[11] Lei H, Zhu XH, Zhang XL, Ugurbil K, Chen W. In vivo 31P magnetic resonance spectroscopy of human brain at 7 T: an initial experience. Magn Reson Med 2003;49(2):199-205.
[12] Du F, Liu S, Li N, Liu C, Chen Q, Zhang S, et al. Performance Evaluation and Comparison of two Triple-nuclear RF Resonators for 1H/19F/31P MR Imaging at 9.4T. *2019 IEEE International Conference on Real-time Computing and Robotics (RCAR).* IEEE; 2019:598-603.
[13] Thulborn KR, Boada FE, Chang SY, Davis D, Gillen JS, Noll DC, et al. Proton, sodium and functional MRI and proton MRS at 1.5 and 3.0 tesla. *ISMRM, 3rd Annual Meeting.* Nice, France; 1995:306.
[14] Sadeghi-Tarakameh A, DelaBarre L, Lagore RL, Torrado-Carvajal A, Wu X, Grant A, et al. In vivo human head MRI at 10.5T: A radiofrequency safety study and preliminary imaging results. Magn Reson Med 2020;84(1):484-96.
[15] Pruessmann KP, Weiger M, Scheidegger MB, Boesiger P. SENSE: sensitivity encoding for fast MRI. Magn Reson Med 1999;42(5):952-62.
[16] Sodickson DK, Manning WJ. Simultaneous acquisition of spatial harmonics (SMASH): fast imaging with radiofrequency coil arrays. Magn Reson Med 1997;38(4):591-603.
[17] Griswold MA, Jakob PM, Heidemann RM, Nittka M, Jellus V, Wang J, et al. Generalized autocalibrating partially parallel acquisitions (GRAPPA). Magn Reson Med 2002;47(6):1202-10.
[18] Lustig M, Donoho, D., & Pauly, J. M. Sparse MRI: The application of compressed sensing for rapid MR imaging. Magnetic resonance in medicine 2007;58(6):1182–95.
[19] Li Y, Chen Q, Wei Z, Zhang L, Tie C, Zhu Y, et al. One-Stop MR Neurovascular Vessel Wall Imaging With a 48-Channel Coil System at 3 T. IEEE Trans Biomed Eng 2020;67(8):2317-27.
[20] Meng Y, Mo Z, Hao Z, Peng Y, Yan H, Mu J, et al. High-resolution intravascular magnetic resonance imaging of the coronary artery wall at 3.0 Tesla: toward evaluation of atherosclerotic plaque vulnerability. Quant Imaging Med Surg 2021;11(11):4522-9.
[21] Tan Y, Chen Q, Xu G, Zhang L, Zhang N, Zhang X, et al. A 48-channel head and neck neurovascular coil for MR vessel wall imaging at 5 T. Proc Intl Soc Mag Reson Med 2023;31:4227.
[22] Li Y, Lafontaine M, Chang S, Nelson SJ. Comparison between Short and Long Echo Time Magnetic Resonance Spectroscopic Imaging at 3T and 7T for Evaluating Brain Metabolites in Patients with Glioma. ACS Chem Neurosci 2018;9(1):130-7.
[23] Connell IR, Gilbert KM, Abou-Khousa MA, Menon RS. Design of a parallel transmit head coil at 7T with magnetic wall distributed filters. IEEE Trans Med Imaging 2015;34(4):836-45.
[24] Rutledge O, Kwak T, Cao P, Zhang X. Design and test of a double-nuclear RF coil for (1)H MRI and (13)C MRSI at 7T. J Magn Reson 2016;267:15-21.
[25] Du F, Liu S, Chen Q, Li N, Dou Y, Yang X, et al. Numerical Simulation and Evaluation of a Four-Channel-by-Four-Channel Double-Nuclear RF Coil for $^1$H MRI and $^{31}$P MRSI at 7 T. IEEE Transactions on Magnetics 2018;54(11):1-5.
[26] Pang Y, Xie Z, Xu D, Kelley DA, Nelson SJ, Vigneron DB, et al. A dual-tuned quadrature volume coil with mixed lambda/2 and lambda/4 microstrip resonators for multinuclear MRSI at 7 T. Magn Reson Imaging 2012;30(2):290-8.
[27] Sengupta S, Roebroeck A, Kemper VG, Poser BA, Zimmermann J, Goebel R, et al. A Specialized Multi-Transmit Head Coil for High Resolution fMRI of the Human Visual Cortex at 7T. PLoS One 2016;11(12):e0165418.
[28] Rietsch SHG, Orzada S, Bitz AK, Gratz M, Ladd ME, Quick HH. Parallel transmit capability of various RF transmit elements and arrays at 7T MRI. Magn Reson Med 2018;79(2):1116-26.
[29] Wu B, Wang C, Krug R, Kelley DA, Xu D, Pang Y, et al. 7T human spine imaging arrays with adjustable inductive decoupling. IEEE Trans Biomed Eng 2010;57(2):397-403.
[30] Zhao W, Cohen-Adad J, Polimeni JR, Keil B, Guerin B, Setsompop K, et al. Nineteen-channel receive array and four-channel transmit array coil for cervical spinal cord imaging at 7T. Magn Reson Med 2014;72(1):291-300.
[31] Lopez-Rios N, Gilbert KM, Papp D, Cereza G, Foias A, Rangaprakash D, et al. An 8-channel Tx dipole and 20-channel Rx loop coil array for MRI of the cervical spinal cord at 7 Tesla. NMR Biomed 2023;36(11):e5002.
[32] Zhao Y, Payne K, Ying L, Zhang X. A coupled planar RF array for ultrahigh field MR imaging. Proc Intl Soc Mag Reson Med 2023;31:3910.